\documentclass[]{spie}  
\usepackage[]{graphicx}
\usepackage{txfonts}
\usepackage{graphicx}
\usepackage{subfigure}
\usepackage{multirow}
\usepackage{rotating}
\usepackage{url}

\title{New Exoplanet Surveys in the Canadian High Arctic at 80 Degrees North}

\author{Nicholas M. Law\supit{a}, Suresh Sivanandam\supit{a}, Richard Murowinski\supit{b}, Raymond Carlberg\supit{c}, Wayne Ngan\supit{c}, Pegah Salbi\supit{a}, Aida Ahmadi\supit{d}, Eric Steinbring\supit{b}, Mark Halman\supit{b}, James Graham\supit{a}
\skiplinehalf
\supit{a}Dunlap Institute for Astronomy and Astrophysics, University of Toronto, 50 St. George Street, Toronto M5S 3H4, Ontario, Canada;
\skiplinehalf
\supit{b}National Science Infrastructure, National Research Council Canada, Victoria, British Columbia, V9E 2E7, Canada
\skiplinehalf
\supit{c}Department of Astronomy and Astrophysics, University of Toronto, 50 St. George Street, Toronto, Ontario M5S 3H4, Canada
\skiplinehalf
\supit{d}University of Calgary, 2500 University Dr. NW, Calgary, Alberta T2N 1N4, Canada
}

\authorinfo{Further author information: send correspondence to law@di.utoronto.ca}

\begin{document} 
\maketitle 

\begin{abstract}

Observations from near the Eureka station on Ellesmere
Island, in the Canadian High Arctic at 80$^{\circ}$ North, benefit from 24-hour darkness combined with
dark skies and long cloud-free periods during the winter.  Our first
astronomical surveys conducted at the site are aimed at transiting
exoplanets; compared to mid-latitude sites, the continuous darkness during the Arctic winter greatly
improves the survey's detection efficiency for longer-period transiting planets. We
detail the design, construction, and testing of the first two instruments: a robotic telescope, and a set of very wide-field
imaging cameras. The 0.5m Dunlap Institute Arctic Telescope has a 0.8-square-degree
field of view and is designed to search for potentially habitable
exoplanets around low-mass stars. The very wide field cameras have several-hundred-square-degree fields of view pointed at Polaris, are designed to search for transiting planets around bright stars, and were tested at the site in February 2012. Finally, we present a conceptual design for the Compound Arctic Telescope Survey (CATS), a multiplexed transient and transit search system which can produce a 10,000-square-degree snapshot image every few minutes throughout the Arctic winter.

\end{abstract}

\section{Introduction}

The continuous wintertime darkness at polar sites can greatly increase the detection efficiency of time-domain astronomy programs. These properties have encouraged the development of Antarctic optical surveys of the Southern sky with imagers and small telescopes\cite{Rico2010, Daban2010}, such as the Gattini cameras\cite{Moore2006,Moore2008,Moore2010} at Dome C ($75^{\circ}$S) and Dome A ($80^{\circ}$S), and the Chinese Small Telescope Array (CSTAR\cite{Wang2011}) which performed long-term photometry on 10,000 stars in a 23 deg$^2$ region centered on the South Celestial Pole. 

In this paper we describe developments towards wide-field photometric monitoring programs in the Canadian High Arctic, which offers 24-hour access to the Northern sky in the winter months. At a latitude of $80^{\circ}$N on Ellesmere Island, Canada's Eureka research base supports the nearby Polar Environment Atmospheric Research Laboratory (PEARL), a facility situated on a 600-m high ridge and designed primarily for atmospheric studies. This ``Ridge Lab'' has been demonstrated to have excellent weather in the winter months\footnote{clear enough conditions for differential photometry $\sim$85\% of the time, and photometric conditions $\sim$50\% of the winter\cite{Steinbring2012}}, good seeing\cite{Hickson2010}, and dark skies\footnote{papers in these proceedings}.

\begin{sidewaysfigure}[tbp]
  \centering
  \resizebox{\textwidth}{!}
   {
	\includegraphics{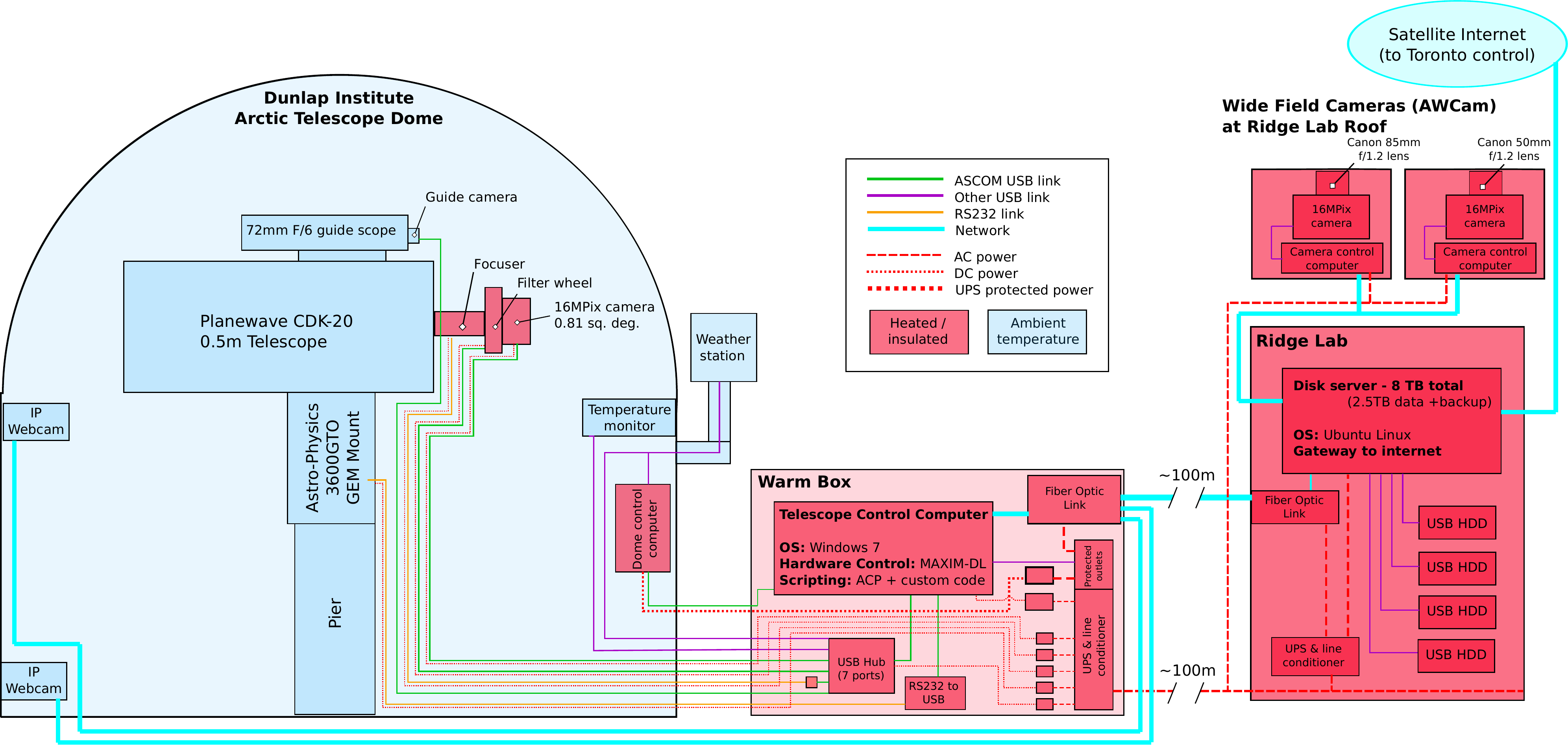}
   }
  \caption{A block diagram of the autonomous wide-field survey facility at the Ridge Lab site. The AWCams have been deployed on the Ridge Lab roof and have demonstrated remote astronomical observatory operation via the Internet. The areas shaded in red are heated and/or insulated; blue areas are allowed to operate at ambient temperature. The "warm box" allows separation of the dome and control electronics from the Ridge Lab building by up to several hundred meters.\label{fig:block_diagram}}
\end{sidewaysfigure}

Our Arctic science instruments are based on two tracks: very-wide-field, small aperture monitoring for both bright exoplanet transit and transient events, and a larger survey telescope designed to search for transiting exoplanets in the habitable zone of $\approx$10,000 M-dwarfs.  Figure \ref{fig:block_diagram} details the major astronomical components installed at or planned for operation at the Ridge Lab site in the near future. The Ridge Lab has proved to be a remarkably good platform for astronomy, providing a base of operations, power, and connectivity to the South via the broadband link of the Canadian Network for the Detection of Atmospheric Change (CANDAC). The wide-field cameras have already been deployed to the site, providing an initial demonstration of reliable operation. The 0.5m Dunlap Institute Arctic Telescope (DIAT) is currently undergoing on-sky testing in New Mexico, along with ongoing cold-testing of the hardware components. In Table \ref{tab:camera_specs} we outline the properties of the systems described in this paper, including a conceptual design for a much larger survey which would build on experience with the current systems.

The challenging environment at the Ridge Lab site requires careful attention to the survivability of equipment deployed there. Instruments must be capable of surviving -50$^{\circ}$C temperatures and storm events depositing snow and ice with wind speeds as high as 40m/s. In general, the snow-accumulation events require manual intervention (a technician with a shovel and broom) to clear instruments and buildings, but the instruments we describe here are otherwise designed to operate autonomously. We note some specific ruggedization strategies for each instrument below.

The paper is organized as follows: in Section \ref{sec:wf_cameras} we evaluate the performance of two wide-field cameras which we deployed at the Ridge Lab site in 2012; in Section \ref{sec:diat} we describe the design and testing of a wide-field, 0.5m telescope which we have ruggedized for Arctic operations; and in Section \ref{sec:cats} we conclude by presenting a conceptual design of a much larger extremely-wide-field, high-cadence sky survey instrument.

\begin{table}
\label{tab:camera_specs}

\begin{small}
\centering
\begin{tabular}{llll}
 & {\bf AWCam systems (deployed)} & {\bf DIAT (under testing)} & {\bf CATS (conceptual)} \\ 
\hline
Field of view & 25.4$^{\circ}$ \& 40.8$^{\circ}$ & 0.93$^{\circ}$ & 120$^{\circ}$\\
Apertures     & 71 mm \& 42 mm & 0.5m & 71 mm\\
Targets       & Exoplanet transits of bright stars & Exoplanet transits of cool stars & Very-high-cadence sky survey \\
Cadence       & 15 seconds & 20 minutes & 1-5 minutes \\
\\
\end{tabular}

\end{small}
\caption{The specifications of the three survey instruments described in this paper.}
\end{table}

\section{Arctic Wide-Field Cameras (AWCams)}
\label{sec:wf_cameras}

\begin{figure}[tb]
  \centering
  \resizebox{0.9\textwidth}{!}
   {
	\includegraphics{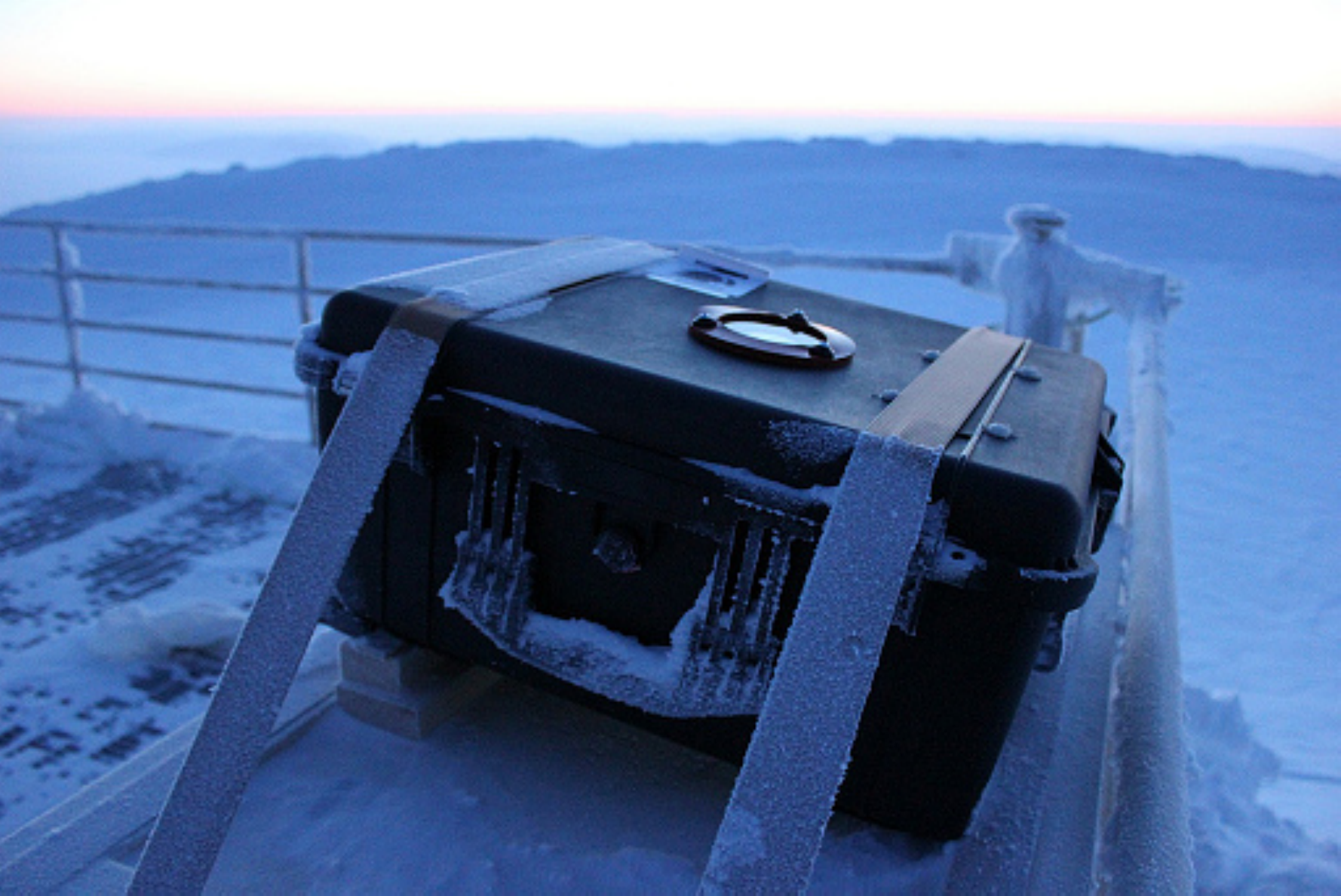}
   }
  \caption{One of the camera enclosures mounted at the Ridge Lab and aligned to the North Celestial Pole. Ice has accumulated around the sides of the case and its surroundings, while the optical window is kept clear of contamination by the heating system. \label{fig:frost}}
\end{figure}

\begin{figure}[tb]
  \centering
  \resizebox{0.9\textwidth}{!}
   {
	\includegraphics{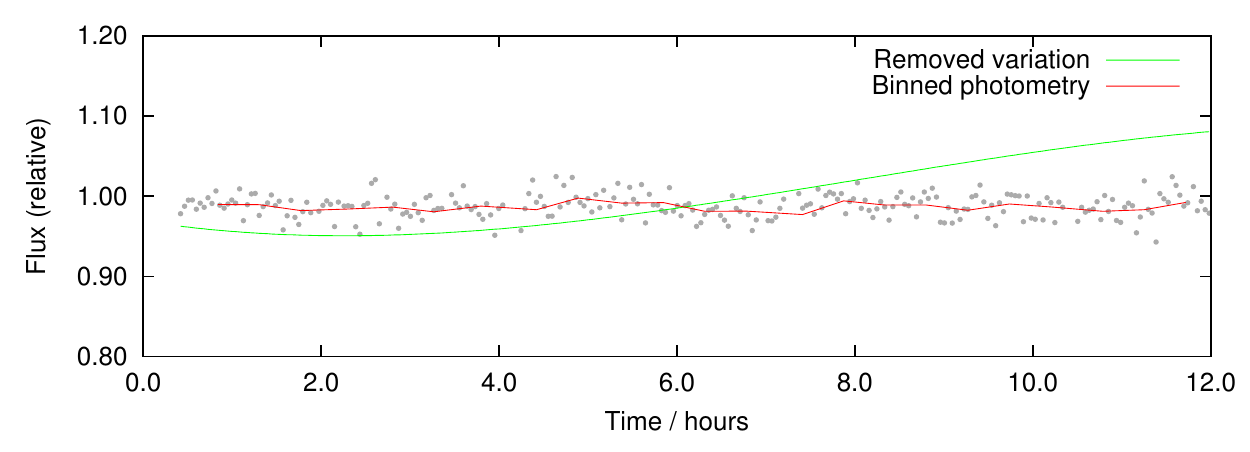}
   }
  \caption{The photometric performance achieved by the 85mm AWCam camera operating with a z-filter, for a typical 12-hour period during our February 2012 arctic observations. The grey points show the individual measured datapoints on a $\rm{m_V\approx9}$ star, after differential calibration against three nearby reference stars of similar brightness. A third-order polynomial (green line) has been subtracted from the measured photometry. The scatter in the individual datapoints is consistent with the expected scintillation-noise limit in these short exposures\cite{Law2012_arctic}. The red line shows bins of 20 datapoints, with a photometric variation of approximately 5 millimagnitudes; increased binning improves the photometric performance to 2-3 millimagnitudes\cite{Law2012_arctic}.  \label{fig:phot}}
\end{figure}

\begin{figure}
  \centering
  \resizebox{1.025\columnwidth}{!}
   {
	\includegraphics{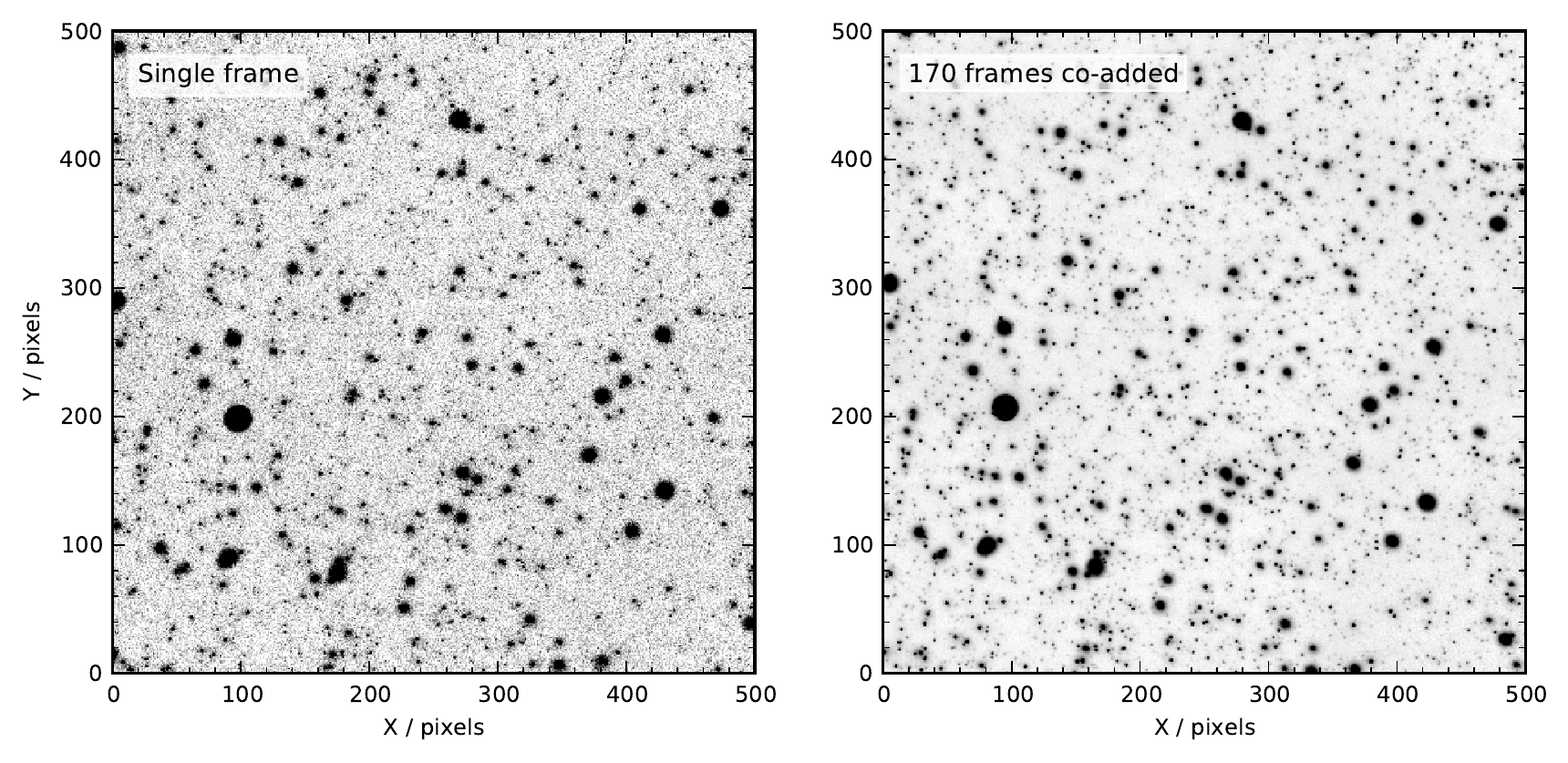}
   }
   \caption{Co-addition of 30 minutes of images: 170 10-second frames from the 85mm camera in the g filter. Co-addition with these systems is challenging because of the large geometric distortions of our lens, the continual field rotation, and the common presence of satellite trails in our images. After astrometrically calibrating our images using the AWCam standard pipeline\cite{Law2012_arctic}, we tested the co-addition performance using SWarp\cite{Bertin2002}. The image resolution is essentially unaffected by the co-addition process, while the imaging depth is greatly increased.}
   \label{fig:coadd}
\end{figure}

As an initial step in precision wide-field photometry, we deployed two wide-field camera systems (AWCams) to the Ridge Lab site in February 2012. We briefly describe the cameras and their performance here; another paper describes and evaluates the systems in much more detail\cite{Law2012_arctic}. 

The AWCam instruments are based on commercial single-lens-reflex camera lenses and large-area CCDs (similar to lower-latitude transiting exoplanet surveys such as SuperWASP\cite{Pollacco2006}, HAT\cite{Bakos2004}, and KELT\cite{Pepper2007, Pepper2012}).. Depending on the lens choice, the optical systems can provide fields of view of hundreds of square degrees with pixel sampling of tens of arcseconds. The systems are designed to detect transiting exoplanets around thousands of bright stars ($\rm m_V = 6-10$). Pointed at the North Celestial Pole, the cameras take short 10-second exposures to avoid star trails induced by sky rotation. The cameras are protected from the Arctic conditions inside modified weather-sealed packing crates, and a heating and air circulation system warms the camera electronics and keeps the optical surfaces clear of snow and ice (Figure \ref{fig:frost}). 

The February 2012 AWCam systems were based on Canon f/1.2 lenses with 85mm and 50mm focal lengths, providing fields of view of 25.4$^{\circ}$ and 40.8$^{\circ}$ respectively. We tested the systems for a total of 152 hours, during which 44,000 images were recorded. One camera system swapped between five filters throughout the observation period, while the other continuously recorded high-cadence in a SDSS-r filter.

The cameras produced 1-4 pixel FWHMs across their fields of view and achieved their photometric performance goals, reaching milli-magnitude photometric precisions (Figure \ref{fig:phot}). High-quality co-addition of the short exposure images allows the cameras to search for faint transient events (Figure \ref{fig:coadd}).

Performance of the initial AWCam instruments was very encouraging. We are confident in reliable autonomous operation throughout a full winter, and are also considering a much larger system which builds on their design (Section \ref{sec:cats}).

\section{The Dunlap Institute Arctic Telescope (DIAT)}
\label{sec:diat}

The DIAT is a 0.5m wide-field imaging telescope designed to search for transiting planets in the habitable zone of cool stars\cite{Law2012}. The telescope is designed to target $\approx$10,000 M-dwarfs at a cadence of $\approx$30-minutes. Approximately 500 M-dwarfs are targeted in each survey field, and several sets of target fields are planned to be imaged throughout the Arctic winter. By taking advantage of the arctic transit detection efficiency increase for longer periods, the system is capable of pushing to detections of planets in the target's habitable zones, which have orbital periods of tens of days. The relatively large signals provided by exoplanets transiting small M-dwarfs increase the detection probability of small planets around these faint targets. The survey's brightness limit is set at a level where radial velocity followup of candidates is feasible with current instruments such as Keck/HIRES\cite{Vogt1994}.  The DIAT is currently performing robotic operation tests and science observations at the New Mexico Skies observatory at Cloudcroft, NM (Figure \ref{fig:diat_pics}).

\subsection{Telescope Hardware}
DIAT's Optical Tube Assembly (OTA) is a PlaneWave Instruments CDK-20 with a corrected-Dall-Kirkham design. The telescope provides an f/6.8 beam which further optics reduce to f/4.5 with an $\approx$1-degree field of view. A 16MPix Apogee U16M camera with 0.82" pixels images a 0.93$^{\circ}$$\times$0.93$^{\circ}$ field.  The telescope is mounted on an Astro-Physics 3600GTO German Equatorial Mount. The telescope is protected by a Technical Innovations 15-ft ProDome, a design which has been tested in cold temperatures and high winds.

\subsection{Software and Robotization}
The telescope hardware control is performed with a mixture of custom and commercial software. Hardware driver support is provided by ASCOM and the hardware manufacturer's drivers; Maxim-DL is used for CCD-camera control; and the ACP Observatory Control Software is used for scripted observations, dome control and weather safety. The use of commercial software allowed the low-level telescope control system to be completed in a very short timescale.

Because a remote site requires highly reliable autonomous operation, hardware safety precautions, and detailed logging, error reporting and hardware monitoring, we implemented custom supervisory software for robotic control. The Python-based software implements a limited queue-scheduling system, automatic dome opening and closing according to weather conditions and science plans, and detailed logging and status reporting via both email and a low-bandwidth communications system (Twitter)\footnote{the telescope currently thoroughly entertains more than a dozen Twitter followers.}.

\subsection{Ruggedizing and Testing for Arctic Conditions}

\begin{figure}[tb]
  \centering
  
  \resizebox{0.7\textwidth}{!}
   {
	\subfigure{\resizebox{0.49\textwidth}{!}{{\includegraphics{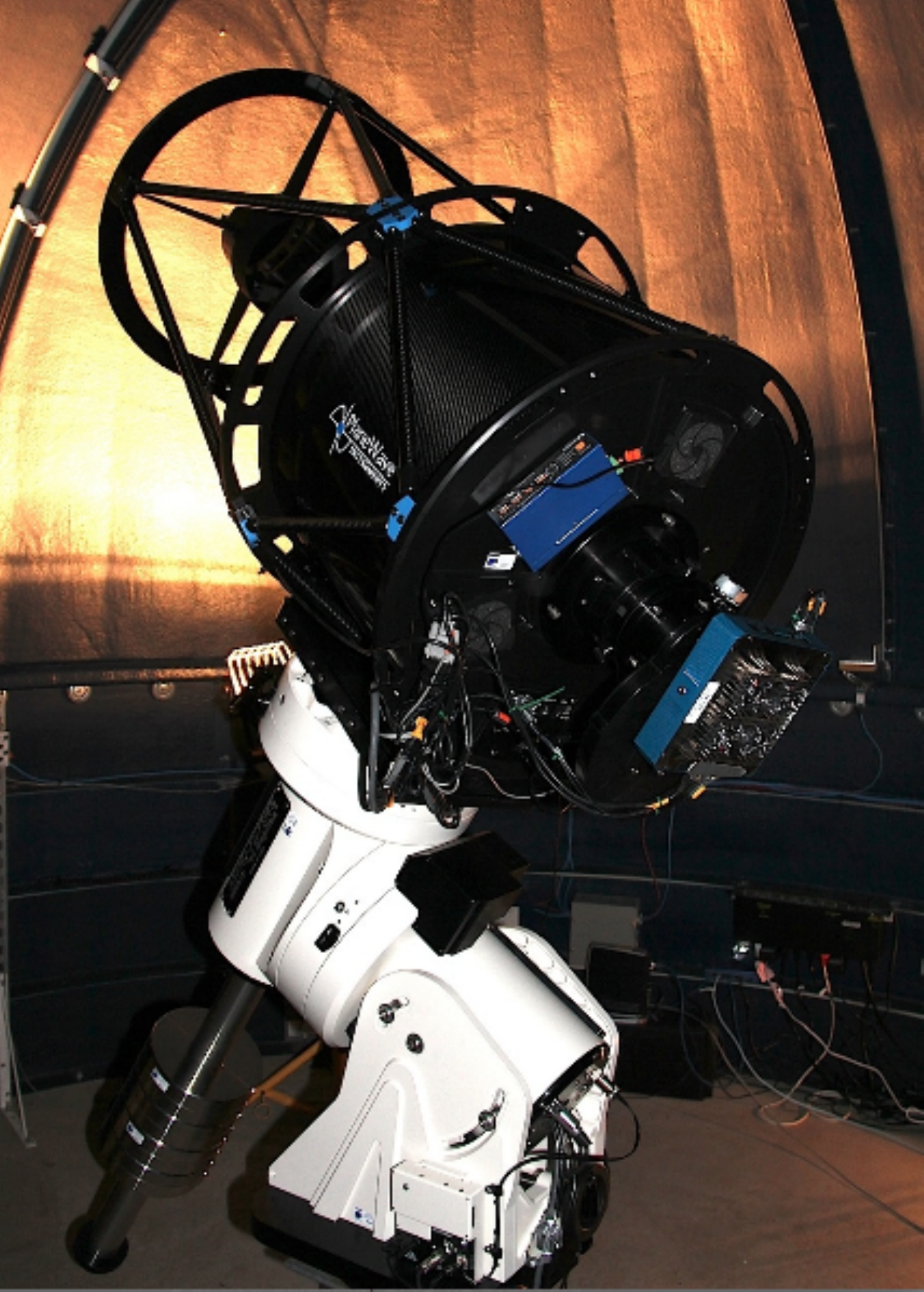}}}}
   	\subfigure{\resizebox{0.486\textwidth}{!}{{\includegraphics{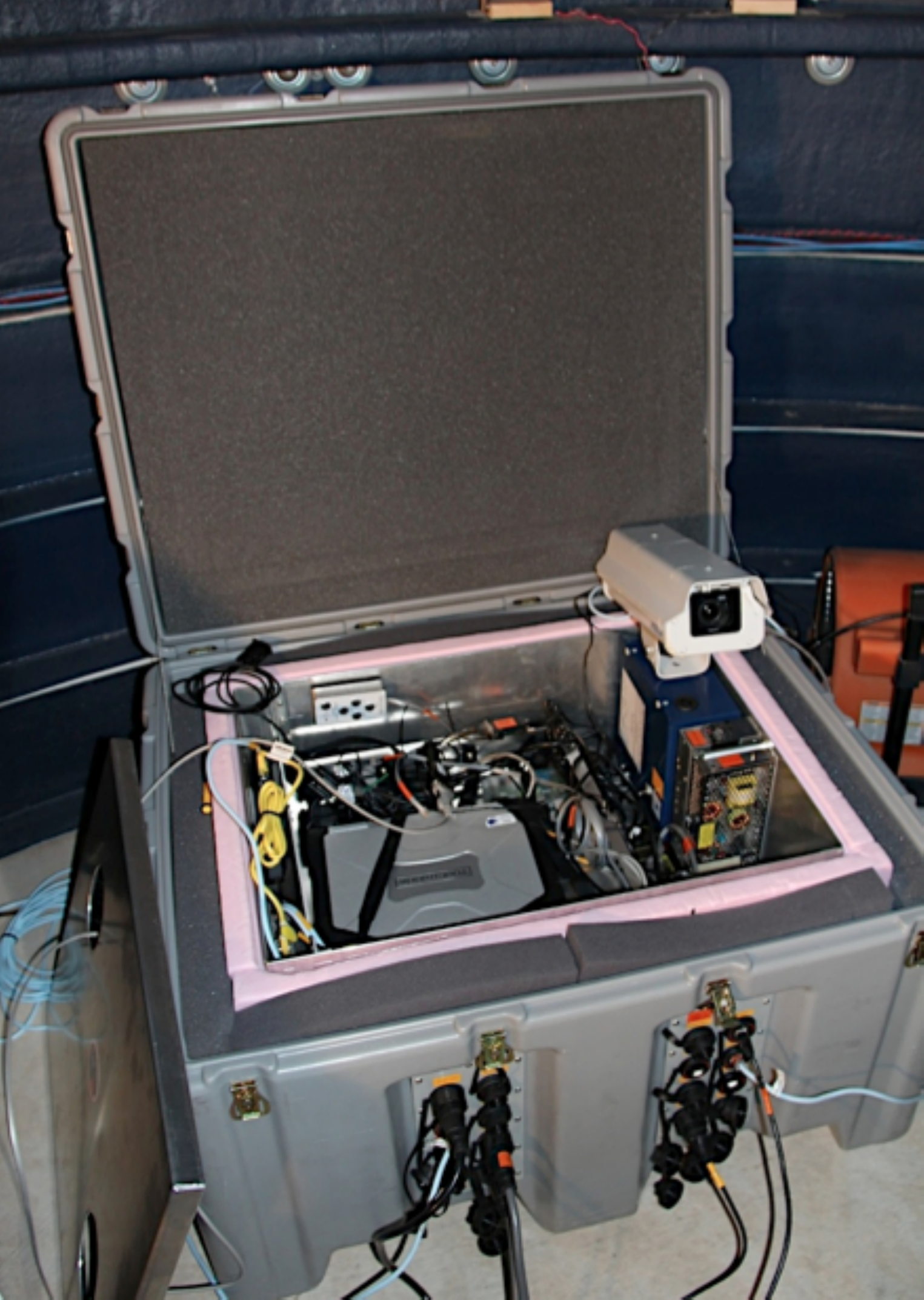}}}}
   }
  \caption{\textit{Left:} The Dunlap Institute Arctic Telescope undergoing testing in New Mexico. \textit{Right:} The custom-built weather-sealed enclosure (shown opened) designed to protect the telescope and dome's control electronics and power supplies from the arctic conditions.\label{fig:diat_pics}}
\end{figure}

The Astro-Physics 3600GTO telescope mount is designed for cold operation when run with a low-temperature grease, although increased wear of the drive train is expected at the lowest temperatures encountered at the Ridge Lab site. The telescope optical elements are likely to be susceptible to ice build-up from "diamond-dust", although the design of the OTA protects all surfaces within baffles, except the secondary mirror. The telescope field correction lenses provide a sealed tube between their location within the primary central hole and the CCD, which reduces contamination possibilities. The exposed optical surfaces will require cleaning of ice accumulation; we are currently testing methods to do this, including remotely-controlled systems (heated blown air).

All the electronic components of the telescope are either designed to operate in the cold, or are protected in heated enclosures. The bulk of our electronics are stored within a custom-built "warm box", an insulated and weather-sealed container (Figure \ref{fig:diat_pics}). 

We tested the telescope's optical quality and mechanical and electronic operation below -30$^{\circ}$C in an NRC cold test chamber. To test the optical quality, we constructed an artificial collimated light source which illuminated the full telescope pupil. Since only a spherical mirror of sufficient size was available, we used it along with a pinhole light source at its focus to construct the collimated beam. The use of the spherical mirror introduced some spherical aberration ($\approx$3/4 wave), but the image quality at the telescope focal plane was sufficient to test for significant optical aberrations introduced due to thermal stresses on the optics. 

An added complication was that the collimator setup was sufficiently large that it had to be housed outside of the cold test chamber. This meant that a window had to be constructed in the cold chamber so the beam can reach the telescope inside. The window could not frost up under normal operations, and double-pane glass windows were not sufficient. A custom window was made out of four sheets of Turbo Film with dry nitrogen filling the air space between them. The window is thought to introduce approximately $\sim$70\,nm RMS of wavefront error, which was acceptable for our work. The window was functional at cold temperatures, though some portions of the film membrane touched each other, which most likely increased the wavefront error. The pupil images no longer showed uniform illumination and the areas of non-uniformity were highly correlated with the areas where the membranes touched each other. Tests at -33C showed a slight degradation of the telescope PSF, which was most likely attributed to poor performance of the Turbo Film window at low temperatures. Overall, there was no evidence for significant degradation of the telescope's optical performance on-axis, and the focus shift was remarkably small once the telescope had equilibriated to the ambient temperatures.

Other electronic equipment including the filter wheel, electronics warm box, focuser and dome control electronics were tested in the cold, and all of them remained functional in simulated Arctic conditions.

\section{The Compound Arctic Telescope Survey (CATS) Conceptual Design}
\label{sec:cats}

\begin{figure}[tb]
  \centering
  
  \resizebox{\textwidth}{!}
   {
	\subfigure{\resizebox{0.57\textwidth}{!}{{\includegraphics{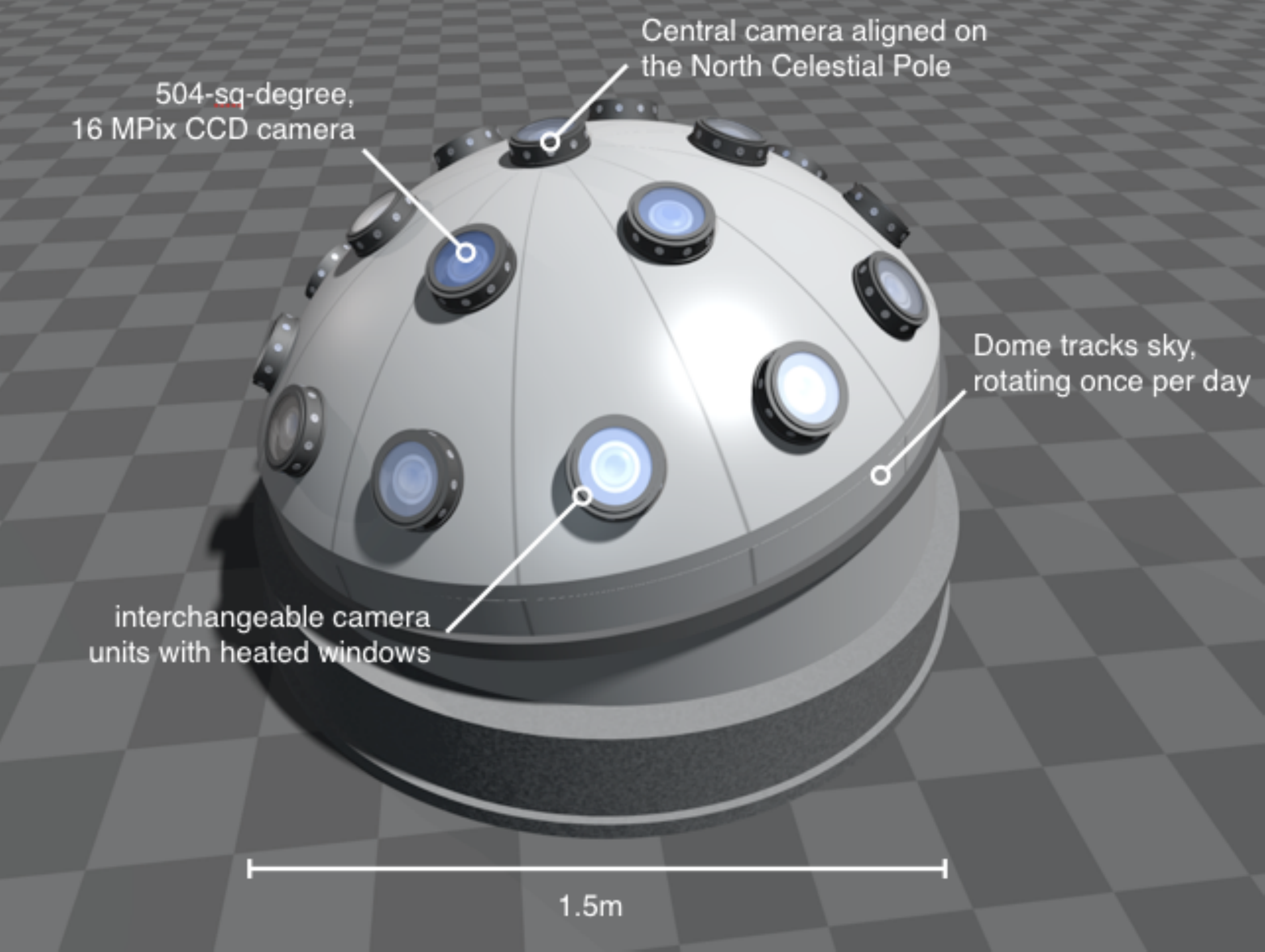}}}}
   	\subfigure{\resizebox{0.43\textwidth}{!}{{\includegraphics{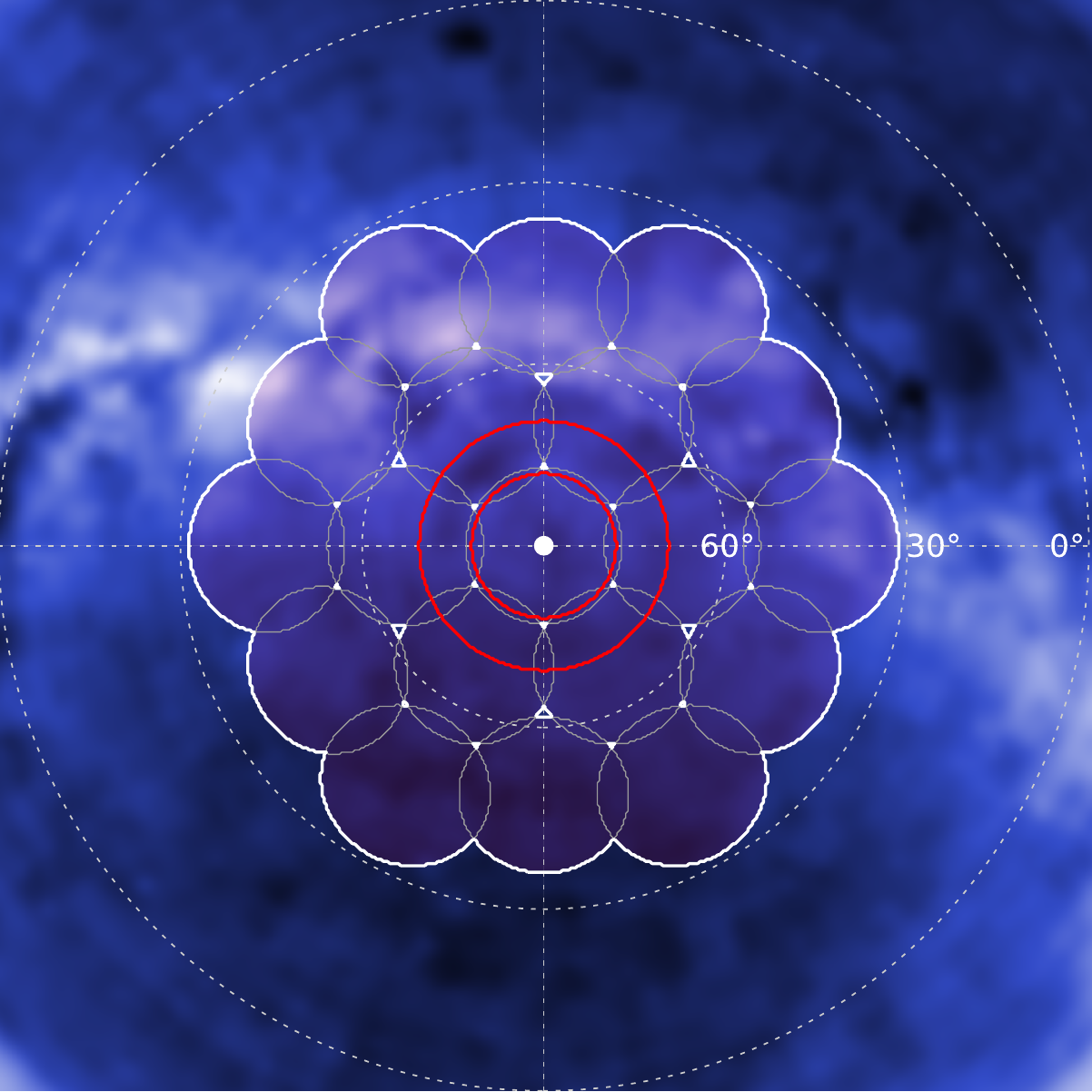}}}}
   }
  \caption{\textit{Left:} A concept drawing for the Compound Arctic Camera System (CATS). Each heated window covers a camera similar to those described in section \ref{sec:wf_cameras}. \textit{Right:} The Northern sky, shown in a polar projection centered on the North Celestial Pole. The fields of view of the two test camera systems are outlined in red. Logarithmically-scaled star counts from the USNO-B1 survey\cite{Monet2003} show the position of the galactic plane. A possible tiling arrangement for the compound arctic telescope survey (CATS) is outlined in white, with the individual cameras shown with grey lines. 19 25-degree field-of-view cameras are arranged in a hexagonal pattern which provides 8600 sq. deg of sky coverage, a 10\%  overlap between adjacent fields for calibration purposes, and an arrangement which is mechanically straightforward and can be made very reliable.\label{fig:cats}}
\end{figure}

The individual wide-field cameras described in section \ref{sec:wf_cameras} cover only 3-6\% of the sky continuously accessible from the Ridge Lab in winter months. To search for rare events in a much larger sky area, we have developed a conceptual design for a compound telescope system consisting of a large number of wide-field cameras mounted in a common enclosure. CATS places the cameras in a hemisphere (Figure \ref{fig:cats}) which rotates with the sky; the circumpolar observing location allows continuous sky tracking for each camera with this mechanically simple and robust one-moving-part system.

The rotating-hemisphere concept allows flexible placement of the camera systems ("unit cameras"). In Figure \ref{fig:cats} we delineate a possible field-of-view for a CATS system based on the circular high-quality PSF field of view of the 85mm f/1.2 lenses used in one of the AWCam systems. The 19 cameras are placed in a hexagonal grid on the sky, reaching declinations as low as 30$^{\circ}$. The arrangement provides excellent packing of the fields of view, while giving a small overlap between adjacent cameras which will facilitate calibration.  The cameras are designed to be interchangeable for both ease of construction and for reliability. However, where required by the science goals, different lenses and filters can be used in different camera units, for example using higher-magnification lenses in more crowded areas.

Using the 85mm f/1.2 lenses on all cameras, the CATS design would cover a total of 8600 square degrees in each snapshot image, including near-complete coverage of the Polar sky down to $\sim40^{\circ}$ degrees declination. The system will be capable of monitoring 90,000 stars brighter than $\rm{10^{th}}$ magnitude, with few-milli-magnitude precision in each few-minute exposure. A further $\sim10^6$ stars can be covered with few-millimagnitude precision with a 30-minute cadence, and stars brighter than $\rm m_V\approx17$ would be monitored with SNR$>5$ every few minutes.

CATS can provide few-minute cadence images for 1/4 of the sky, and co-adding of its images (Section \ref{sec:wf_cameras}) allows faint objects to be detected. However, its relatively large pixel size leads to source crowding at fainter magnitudes, and means the system is best used for objects brighter than $\rm{m_V} < 18$ in typical sky regions. CATS would be particularly productive searching for variability of relatively bright sources, such as in exoplanet searches, variable stars and the brightest and most rapid extragalactic transients.

The recent detections of transiting planets around very bright stars\cite{Charbonneau2000, Henry2000, Sato2005, Bouchy2005, Christian2006, Bakos2007, Pal2010, Winn2011, Howell2012}, and the relative ease of characterization of those planets\cite{Agol2010, Collier2010, Demory2011, vonBraun2011, Stevenson2011, Majeau2012}, have demonstrated the effectiveness of using small-aperture telescopes to search for exoplanets. With the photometric precision already demonstrated by our AWCam systems, extrasolar planets as small as Neptune could be found around at least 90,000 very bright stars in the CATS survey area. Bright, nearby star-star microlensing events will occur within the CATS field\cite{Han2008}. The few-minute temporal resolution, high photometric precision and 24-hour coverage of CATS would potentially allow the detection of planetary microlensing events in the discovery light curves, along with providing very rapid alerts about interesting events. At extragalactic distances, bright nearby supernovae, optical gamma-ray-burst afterglows, and other transients would be found within minutes of the objects reaching a detectable brightness anywhere in the system's near-10,000 square degree sky coverage area.

\section{Summary}

The Canadian High Arctic offers an opportunity for 24-hour access to a dark Northern sky, with excellent weather conditions and potentially very good seeing. Astronomical science operations with wide-field telescopes began in February 2012, and they are planned to continue with full-winter campaigns and the addition of the DIAT 0.5m robotic telescope. The CATS conceptual design could provide continuous imaging coverage of much of the Northern sky; coupled to an upgraded Dunlap Institute Arctic Telescope equipped with a low-resolution spectrograph, the system would provide an integrated, rapid High Arctic sky survey and follow-up facility.

\acknowledgments          

We thank Liviu Ivanescu and Paul Hickson for very useful discussions, and Gordon Walker, Russel Robb, Dmitry Monin, Peter Byrnes, Murray Fletcher and Brian Leckie for participation in a planning review of the DIAT. It is our pleasure to acknowledge the arctic expertise of Pierre Fogal and James Drummond from CANDAC, and for their help during operations at the PEARL. We are also indebted to Environment Canada and the staff of the Eureka weatherstation for their hospitality and support of our observing runs. N.M.L. and S.S. are supported by Dunlap Fellowships at University of Toronto. This project was partially supported by funds from the Natural Sciences and Engineering Research Council of Canada, and the National Research Council of Canada. The research made use of tools provided by Astrometry.net and NASA's Astrophysics Data System Bibliographic Services.


\bibliography{refs}   
\bibliographystyle{spiebib}   
\end{document}